\documentclass{ws-p8-50x6-00}
\usepackage{graphicx}
\usepackage{amsmath}

\begin{document}

\pagestyle{empty}

\begin{flushleft}
{RIKEN-AF-NP-402, SAGA-HE-177-01, TMU-NT-01-05 \hfill June 19, 2001} \\
\end{flushleft}
\vspace{1.8cm}
 
\begin{center}
\LARGE{\bf Parton distribution functions in nuclei} \\
\vspace{1.5cm}
\Large{M. Hirai,$^1$ S. Kumano,$^2$ and M. Miyama\,$^3$ $^*$}  \\
\vspace{0.5cm}
\large{$^1$ Radiation Laboratory, RIKEN (The Institute of Physical and \\
            Chemical Research), Wako, 351-0198, Japan} \\
\vspace{0.2cm}
{$^2$ Department of Physics, Saga University,
      Saga, 840-8502, Japan} \\
\vspace{0.2cm}
{$^3$ Department of Physics, Tokyo Metropolitan University \\
         Tokyo, 192-0397, Japan} \\
\vspace{1.5cm}
\Large{Talk given at the 9th International Workshop on} \\
{Deep Inelastic Scattering} \\
{Bologna, Italy, April 27 $-$ May 1, 2001} \\
{(talk on April 28, 2001) }  \\
\end{center}
\vspace{1.6cm}
\noindent{\rule{6.0cm}{0.1mm}} \\
\vspace{-0.3cm}
\normalsize

\noindent
{* Email: mhirai@rarfaxp.riken.go.jp, kumanos@cc.saga-u.ac.jp,} \\
\noindent
{\ \, miyama@comp.metro-u.ac.jp. Information on their research
        is available at} \\
\noindent
{\ \, http://hs.phys.saga-u.ac.jp.}  \\

\vspace{+0.1cm}
\hfill {\large to be published in proceedings}
\vfill\eject
\setcounter{page}{1}
\pagestyle{plain}


\title{Parton distribution functions in nuclei}

\author{M. Hirai,$^1$ S. Kumano,$^2$ M. Miyama$^3$}

\address{$^1$ Radiation Laboratory, RIKEN (The Institute of Physical and
              Chemical Research) \\ Wako, 351-0198, Japan \\
         $^2$ Department of Physics, Saga University,
              Saga, 840-8502, Japan \\
         $^3$ Department of Physics, Tokyo Metropolitan University,
              Tokyo, 192-0397, Japan}

\maketitle

\abstracts{Optimum nuclear parton distributions are determined by an analysis
           of muon and electron deep inelastic scattering data. 
           Assuming simple $A$ dependence and polynomial functions of $x$
           and $1-x$ for nuclear modification of parton distributions,
           we determine the initial distributions by a $\chi^2$ analysis.
           Although valence-quark distributions are relatively well
           determined except for the small-$x$ region, antiquark
           distributions cannot be fixed at medium and large $x$. 
           It is also difficult to fix gluon distributions.}

\section{Introduction}
Although parton distribution functions in nuclei are often assumed to
be equal to those in the nucleon, it is especially important to know
the details of nuclear modification in order to find any exotic signature
in hadron reactions. For example, we discuss a reaction such as 
$J/\psi$ production, which is very sensitive to nuclear gluon distributions,
as a possible way to find a quark-gluon plasma signature. However, it is
unfortunate that the gluon modification is essentially unknown at this stage
although there are some implications from lepton deep inelastic data.

There were recent trials to obtain nuclear parton distributions from
experimental data, for example, by Eskola, Kolhinen, and Ruuskanen.\cite{ekr}
Their studies are valuable for providing possible nuclear modification 
from available data. However, the distributions should be optimized
in principle by a $\chi^2$ analysis. Because there are not so many available
experimental data in the nuclear case, it is obvious that such an effort
is not an easy work at this stage. In Ref. 2, a possible method is developed
as a first step trial for the nuclear $\chi^2$ analysis.
This talk is based on this work. Our analysis method and
results are explained in the following.

\section{Analysis method}

An important point in the analysis is how to set up a functional
form of nuclear parton distributions. Nuclear modification of parton
distributions is typically less than 20\% for medium size nuclei, so that
it is more appropriate to parametrize the modification instead
of the distributions themselves. The nuclear parton distributions are
then assumed as
\begin{equation}
f_i^A (x, Q_0^2) = w_i(x,A,Z) \, f_i (x, Q_0^2),
\label{eqn:apart}
\end{equation}
where $f_i(x, Q_0^2)$ is the $i$-type parton distribution 
($i=u_v$, $d_v$, $\bar q$, $g$) in the nucleon at $Q_0^2$.
The distributions in the nucleon are taken from the MRST
parametrization.\cite{mrst98}
Nuclear antiquark distributions are assumed to be flavor
symmetric: $\bar u^A$=$\bar d^A$=$\bar s^A$ at $Q_0^2$.
We call $w_i(x,A,Z)$ a weight function, which is 
determined by a $\chi^2$ analysis.

Mass-number dependence of $w_i$ could be complicated because different physics
mechanisms contribute depending on the $x$ region. On the other hand, it is
necessary to simplify the problem as a first effort of nuclear $\chi^2$
analysis. Here, we decided to assume the $A$ dependence as $1/A^{1/3}$
as suggested by Sick and Day.\cite{sd}
Then, the rest is taken as a polynomial function of $x$ and $1-x$:
\begin{equation}
w_i(x,A,Z)  = 1+\left( 1 - \frac{1}{A^{1/3}} \right) 
          \frac{a_i(A,Z) +b_i x+c_i x^2 +d_i x^3}{(1-x)^{\beta_i}},
\label{eqn:wi}
\end{equation}
where $a_i$, $b_i$, $c_i$, $d_i$, and $\beta_i$ are parameters to be
determined by the $\chi^2$ analysis. We call this function a ``cubic type"
and the function without the $d_i x^3$ term a ``quadratic type".
Analyses are done for both cases.

There are restrictions on the nuclear parton distributions due to 
the following three conditions:
\begin{alignat}{2}
\text{Charge:} & \ \ \ \  & Z & 
             = \int dx \, \frac{A}{3} \left ( 2 \, u_v^A - d_v^A \right ) ,
\label{eqn:charge}
\\
\text{Baryon Number:} & \ \ \ \  & A & 
             = \int dx  \, \frac{A}{3} \left (  u_v^A + d_v^A  \right ) ,
\label{eqn:baryon}
\\
\text{Momentum:} & \ \ \ \  & A & 
             = \int dx \, A \, x 
            \left ( u_v^A + d_v^A  + 6 \, \bar q^A + g^A \right ) ,
\label{eqn:momentum}
\end{alignat}
where the distributions are defined by the ones per nucleon.
Because of these constraints, three parameters can be expressed
by the others. 
In addition, we fixed some antiquark and gluon parameters which become
relevant at large $x$. Detailed explanations should be found in our
first paper.\cite{saga01}

From the parton distributions in Eq.(\ref{eqn:apart}), the structure
function $F_2$ is calculated by the leading-order expression:
$F_2^A (x,Q^2) = \sum_q e_q^2 x [ q^A(x,Q^2) + \bar q^A(x,Q^2) ]$.
In calculating $F_2^A (x,Q^2)$, the initial distributions at $Q_0^2$
are evolved to $Q^2$ by the ordinary DGLAP evolution equations.
Experimental data are taken from the measurements by the EMC,
BCDMS, NMC, SLAC-E49, -E87, -E139, -E140, and Fermilab-E665
collaborations. The total number of the data is 309.
Various nuclear targets are used experimentally.
In our theoretical analysis, they are assumed as
$^4$He, $^7$Li, $^9$Be, $^{12}$C, $^{14}$N, $^{27}$Al, $^{40}$Ca,
$^{56}$Fe, $^{63}$Cu, $^{107}$Ag, $^{118}$Sn,
$^{131}$Xe, $^{197}$Au, and $^{208}$Pb. 
In comparison with the experimental data, $\chi^2$ is calculated
with the ratio $R_{F_2} ^A = F_2^A / F_2^D$ as
\begin{equation}
\chi^2 = \sum_j \frac{(R_{F_2,j}^{A,data}-R_{F_2,j}^{A,theo})^2}
                     {(\sigma_j^{data})^2}.
\label{eqn:chi2}
\end{equation}

\vspace{-0.6cm}
\section{Results}

\begin{figure}[h]
\parbox[t]{0.46\textwidth}{
   \begin{center}
\includegraphics[width=0.40\textwidth]{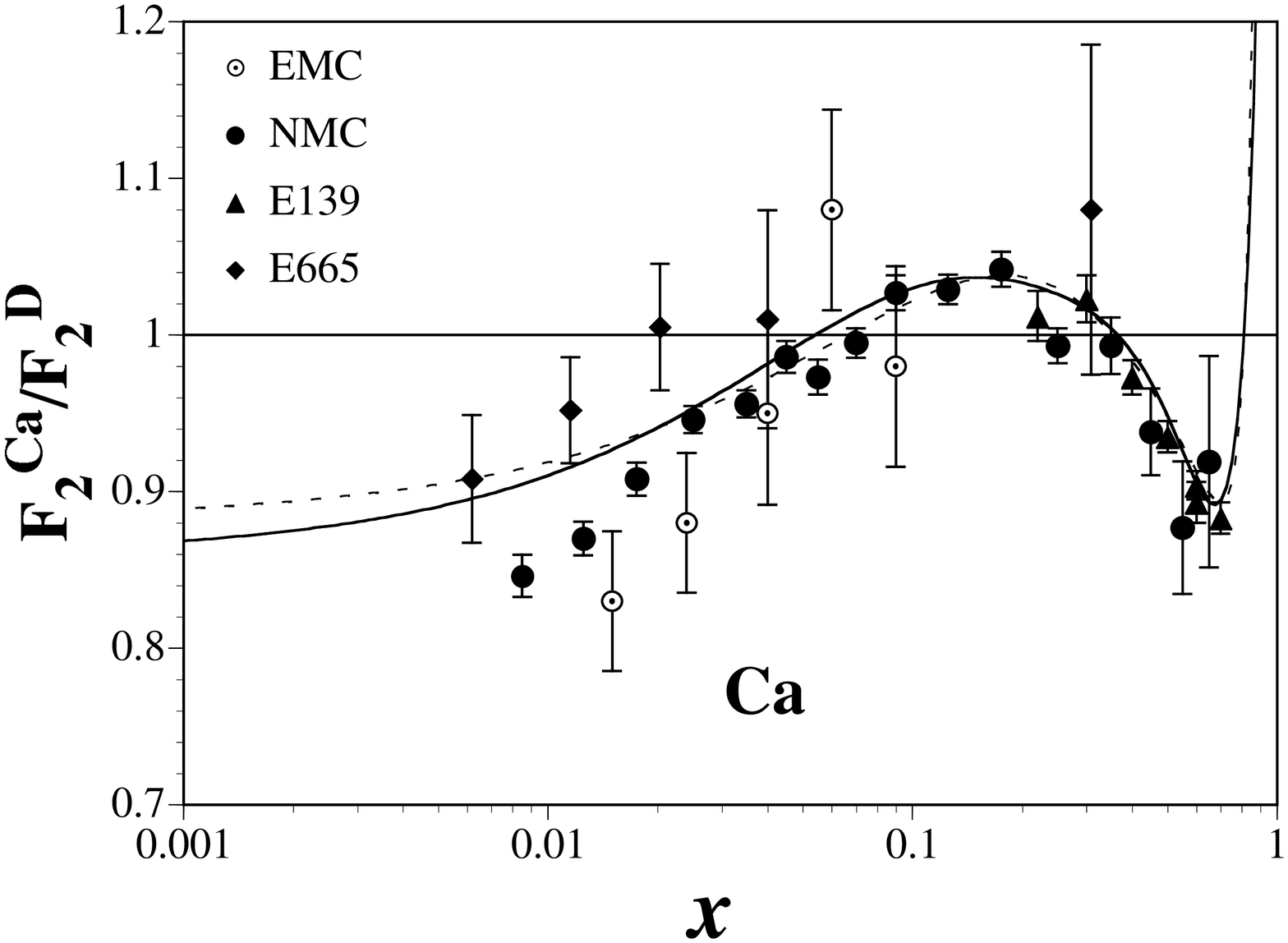}
   \end{center}
   \vspace{-0.5cm}
       \caption{\footnotesize Comparison with calcium data.}
       \label{fig:ca}
}\hfill
\parbox[t]{0.46\textwidth}{
   \begin{center}
\includegraphics[width=0.40\textwidth]{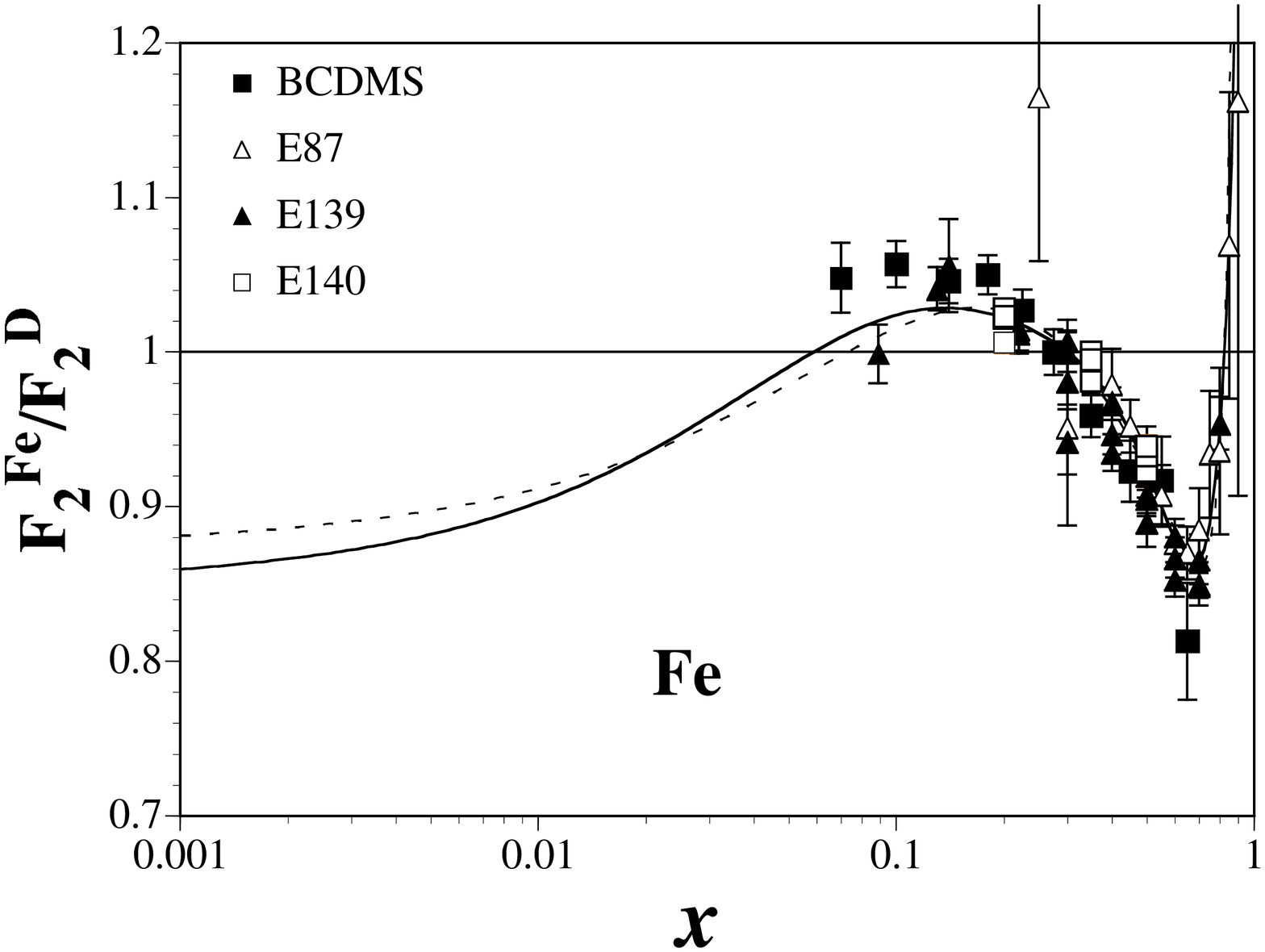}
   \end{center}
   \vspace{-0.5cm}
       \caption{\footnotesize Comparison with iron data.}
       \label{fig:fe}
}
\end{figure}

\begin{figure}[h]
\parbox[t]{0.46\textwidth}{
   \begin{center}
\includegraphics[width=0.40\textwidth]{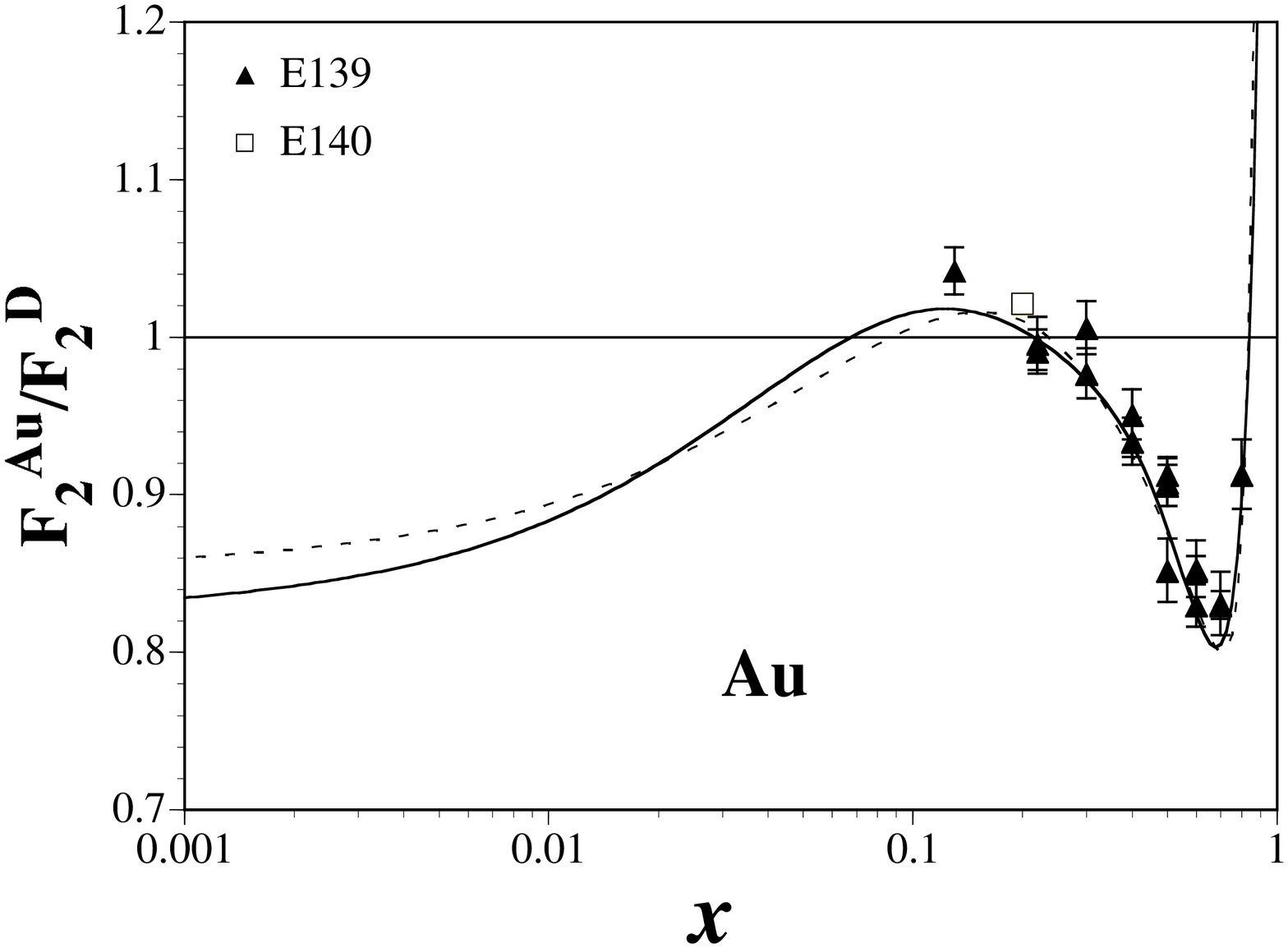}
   \end{center}
   \vspace{-0.4cm}
       \caption{\footnotesize Comparison with gold data.}
       \label{fig:au}
}\hfill
\parbox[t]{0.46\textwidth}{
   \begin{center}
\includegraphics[width=0.40\textwidth]{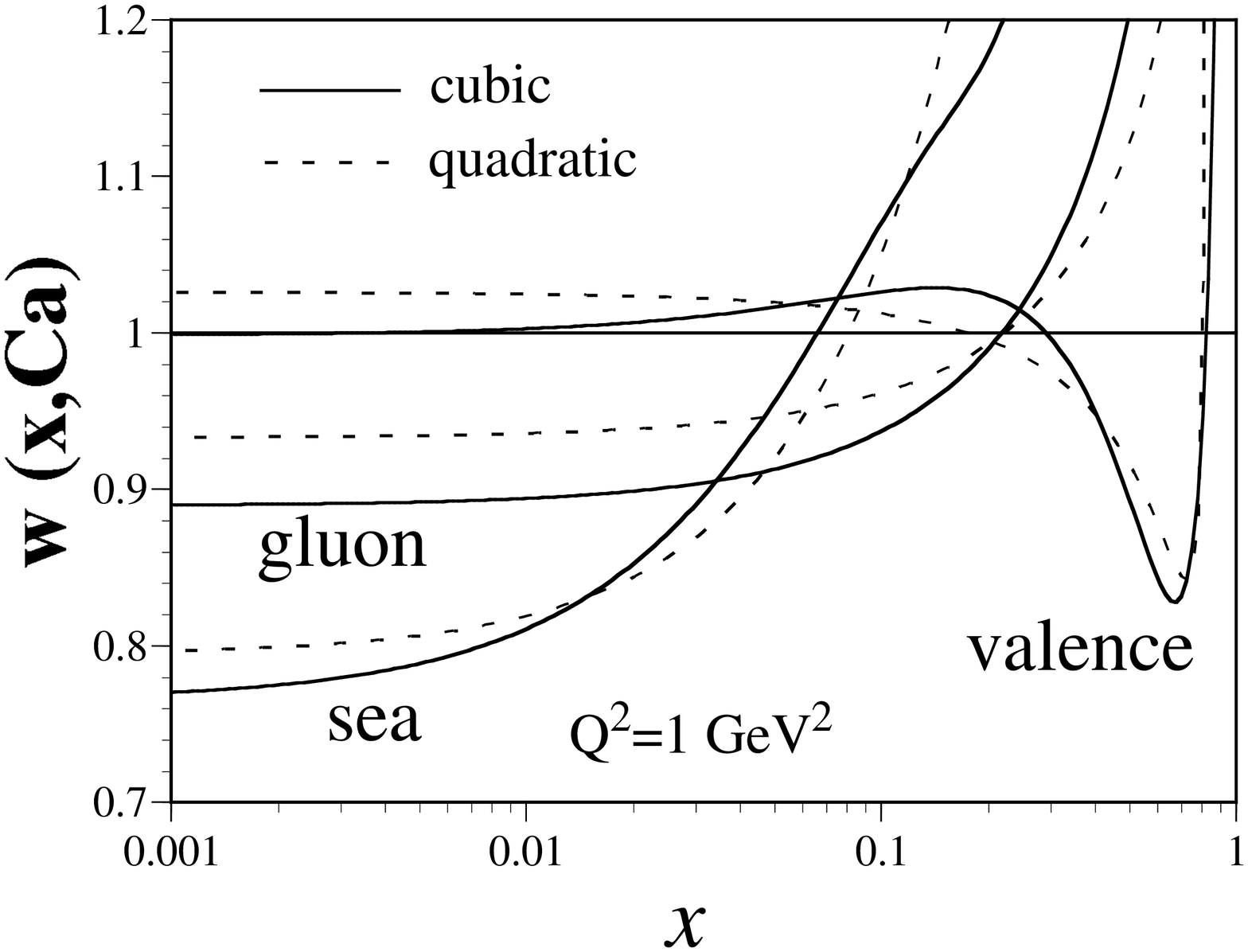}
   \end{center}
   \vspace{-0.2cm}
       \caption{\footnotesize Weight functions in calcium.}
       \label{fig:wca}
}
\end{figure}

Finding the minimum point of $\chi^2$, we determine the parameters.
Obtained distributions are compared with the data in Figs.\ref{fig:ca},
\ref{fig:fe}, and \ref{fig:au}, where the dashed and solid curves
indicate quadratic and cubic results, respectively.
Because the theoretical curves are calculated at
$Q^2$=5 GeV$^2$ and the data are taken at various $Q^2$ points,
they cannot be compared directly. However, the figures seem to
indicate reasonable agreement with the data.
Obtained $\chi^2$ per degrees of freedom is given by
$\chi_{min}^2/d.o.f.$=1.93 in the quadratic fit and 1.82 in the cubic fit.
Because of the additional freedoms, $\chi^2$ is slightly better
in the cubic analysis. Obtained $\chi_{min}^2$ values are not excellent
in the sense $\chi_{min}^2/d.o.f.>1$,
but they are partly due to the scattered experimental data as
obvious in Fig.\ref{fig:ca}. 

\vspace{-0.3cm}
\noindent
\vspace{-0.4cm}
\noindent
\vspace{-0.3cm}

In order to compare both analysis results, we show the weight functions
of the calcium nucleus in Fig.\ref{fig:wca}.
It indicates that the weight functions depend slightly on
the assumed functional form. 
It is noteworthy to mention that the valence-quark distributions
do not show any strong shadowing as $F_2$ or the antiquark distributions.
It is not clear at this stage whether this is an artifact due to 
the lack of Drell-Yan data in our analysis. 
In our studies, we have just set up a method of nuclear $\chi^2$ fit.
In future, we need to improve our method and to include other existing data.
Nuclear parton distributions can be calculated by obtaining
computer codes from our web site.\cite{nucl-lib}
The distributions are provided for nuclei from the deuteron to heavy ones.
In addition, analytical expressions of the weight functions are given
in Appendix of Ref. 2. 

\section{Summary}
Using electron and muon deep inelastic scattering data, we have investigated
optimum parton distributions in nuclei. Valence-quark distributions are
well determined except for the small-$x$ region. Antiquark and gluon
distributions cannot be determined well at medium and large $x$.
The gluon distributions cannot be fixed
because of a leading-order analysis and lack of sensitive data.

\section*{Acknowledgments}
S.K. and M.M. were supported by the Grant-in-Aid for Scientific
Research from the Japanese Ministry of Education, Culture, Sports,
Science, and Technology. M.M. was also supported by a JSPS Research
Fellowship.



\end{document}